\documentclass{acm_proc_article-sp}
\usepackage{times}
\usepackage{graphicx}
\usepackage{psfrag}
\usepackage{enumerate}
\usepackage{amsmath,amssymb,amsthm1}

\newcommand{\hide}[1]{}
\usepackage{colortbl}
\usepackage{fancybox}
\newlength{\mediumwidth}
\newlength{\quotewidth}
\makeatletter
\if@twocolumn
  \AtBeginDocument{%
    \setlength{\mediumwidth}{0.95\columnwidth-\shadowsize-2\fboxsep}
    \setlength{\quotewidth}{0.9\columnwidth}}
\else
  \AtBeginDocument{%
    \setlength{\mediumwidth}{0.8\textwidth}  
    \setlength{\quotewidth}{0.66\textwidth}}
\fi
\makeatother
\newcommand{\framelabel}{}%
\newsavebox{\framelabelbox}%
\newenvironment{frameandlabel}[1]{%
  \vspace{0.6\baselineskip plus 0.5\baselineskip minus 0.2\baselineskip}%
  \renewcommand{\framelabel}{#1}%
  \begin{lrbox}{\framelabelbox}%
    \begin{minipage}{\mediumwidth}%
}{%
    \vspace*{-2mm}
    \end{minipage}%
  \end{lrbox}%
  \begin{center}
    \shadowsize=1pt
    \shadowbox{%
      \begin{minipage}{\mediumwidth}%
        \centerline{\makebox[0pt]{\raisebox{0.5ex}[0pt]{\colorbox{white}{\footnotesize\textsc{\framelabel}}}}}%
        \usebox{\framelabelbox}%
        \vspace*{1ex}%
      \end{minipage}}
  \vspace{-2ex}
  \end{center}%
}

\theoremstyle{plain}
\newtheorem{theorem}{Theorem}
\newtheorem{lemma}[theorem]{Lemma}

\theoremstyle{definition}
\newtheorem{definition}{Definition}
\newtheorem{construction}{Construction}

\theoremstyle{remark}
\newtheorem{remark}{Remark}

\newcommand{\bs}[1]{\boldsymbol{#1}}
\newcommand{\x}{\bs{x}}

\newcommand{\n}{\bs{\nu}}
\newcommand{\brk}[1]{\left( #1 \right)}

\newcommand{\deriv}[2]{\frac{d#1}{d#2}}
\newcommand{\pd}[2]{\frac{\partial#1}{\partial#2}}
\newcommand{\calD}{\mathcal{D}}
\newcommand{\calQ}{\mathcal{Q}}

\begin{document}

\setcounter{page}{0}

\title{No Justified Complaints:\\ On Fair Sharing of Multiple
  Resources}
\numberofauthors{1}
\author{Danny Dolev$^1$ ~~ Dror G. Feitelson$^1$ ~~
  Joseph Y. Halpern$^2$\thanks{Much of this work was done while the
  author was on sabbatical leave at Hebrew University.} ~~
  Raz Kupferman$^3$ ~~ Nati Linial$^1$\\[2mm]
\affaddr{$^1$School of Computer Science and Engineering, Hebrew
  University, Jerusalem, Israel}\\
\affaddr{$^2$Computer Science Dept., Cornell University, Ithaca, NY}\\
\affaddr{$^3$Institute of Mathematics, Hebrew University, Jerusalem,
  Israel}}


\maketitle
\pagenumbering{arabic}
\pagestyle{plain}

\begin{abstract}
  Fair allocation has been studied intensively in both economics and
  computer science, and fair sharing of resources has aroused renewed
  interest with the advent of virtualization and cloud computing.
  Prior work has typically focused on mechanisms for fair sharing
  of a single resource.
  We provide a new definition for the simultaneous fair allocation of
  multiple continuously-divisible resources.
  Roughly speaking, we define fairness as the situation where every
  user either gets all the resources he wishes for, or else gets at
  least his entitlement on some \emph{bottleneck resource}, and therefore
  cannot complain about not getting more.
  This definition has the same desirable properties as the recently
  suggested dominant resource fairness, and also handles the case of
  multiple bottlenecks.
  We then prove that a fair allocation according to this definition is
  guaranteed to exist for any combination of user requests and
  entitlements (where a user's relative use of the different resources
  is fixed).
  The proof, which uses tools from the theory of ordinary differential
  equations, is constructive and provides a method to compute the
  allocations numerically.
\end{abstract}

\category{D.4.1}{OPERATING SYSTEMS}{Process Management}[Scheduling]
\category{K.6.2}{MANAGEMENT OF COMPUTING AND INFORMATION
  SYSTEMS}{Installation Management}[Pricing and resource allocation]

\terms{Management, performance}

\keywords{Resource allocation, fair share, bottlenecks}

\section{Introduction}

\emph{Fair allocation} is a problem that has been widely studied both
in economics and computer science.
In economics, a wide range of issues have been studied, ranging from
the design of voting rules and the apportionment of representation in
Congress to the allocation of joint costs and fair cake cutting to
envy-free auctions.
(See \cite{brams96,goldberg03,young85} for a sample of the
wide-ranging work in the area.)
In computer science, besides the work on economics-related issues,
fair allocation has been the focus of a great deal of attention in
operating systems, where fair-share scheduling is a major concern.
(See the related work in Section~\ref{sect:prev}.)

But what exactly does fair allocation mean?  
Generally speaking, the notion of fairness may pertain to mechanisms
like bargaining and their relationship to ethical issues
(e.g.\ \cite{yaari84}).
We focus on a more technical level, and take ``fair
allocation'' to mean ``an allocation according to agreed
entitlements''.
The source of the entitlements is immaterial;
for example, they could result from unequal contributions towards the
procurement of some shared computational infrastructure, or from the
dictates of different service-level agreements.
Still, what does it mean to say that a user is ``entitled to 20\% of
the system''?
Is this a guarantee for 20\% of the CPU cycles? Or maybe 20\% of each
and every resource?
And what should we do if the user requires, say, only 3\% of
the CPU, but over 70\% of the network bandwidth?
Reserving 20\% of the CPU for it will cause obvious waste, while
curbing its network usage might also be ill-advised if no other user
can take up the slack.

Our goal in this paper is to define a notion of fair allocation when
multiple, continuously-divisible resources need to be allocated, and
show that a fair allocation according to our definition is guaranteed
to exist.
Our motivation comes from work in operating systems, so many of our
examples and much of the discussion below is taken primarily from
that literature.
But, as should be clear, our approach is meaningful whenever a number
of users need to share a number of different resources, each has a
certain pre-negotiated entitlement to a share of the resources, but
each has different needs for each of the resources.

The common approach to resource management in both operating systems
and virtual machine monitors (VMMs) is to focus on the CPU.
Scheduling and allocation are done on the CPU, and this induces a use
of other devices such as the network or disk.
However, the relative use of diverse devices by different
processes may be quite dissimilar.
For example, by trying to promote an I/O-bound process (because it
deserves more of the CPU than it is using), we might turn
the disk into a bottleneck, and inadvertently allow the internal
scheduling of the disk controller to dictate the use of the whole
system.
Thus the CPU-centric view may be inappropriate when the goal is to
achieve a predefined allocation of the resources.

In order to avoid such problems, it has recently been suggested that
fair-share scheduling be done in two steps \cite{etsion09}:
first, identify the resource that is the system
bottleneck, and then enforce 
the desired relative allocations on this resource.
The fair usage of the bottleneck resource induces some level of
usage of other resources as well, but this need not be controlled,
because there is sufficient capacity on those resources for all
contending processes.

The question is what to do if two or more resources become
bottlenecks.
This may easily happen when different processes predominantly use
distinct resources.
For example, consider a situation where one process makes heavy use of
the CPU, a second is I/O-bound, while a third process uses both CPU
and I/O, making both bottlenecks.
We consider such situations and make the following contributions.

We propose a definition of what it means to be fair that is appropriate
even when different users or processes have different requirements for
various resources.
The definition, presented in Section \ref{sect:bottleneck}, extends
the idea of focusing on the bottleneck;
it essentially states that we are fair as long as \emph{each and every
user receives his entitlement on at least one bottleneck resource}.
This is claimed to be fair because, given an assumption that each
user uses the different resources in predefined proportions, the
definition implies that users cannot justifiably complain about not
getting more.
We then prove in Section \ref{sect:exist} that an allocation that
satisfies our fairness definition is guaranteed to exist.
The proof is constructive and provides a method to compute the
allocation numerically.
Perhaps surprisingly, the proof makes use of tools from the theory of
ordinary differential equations.
To the best of our knowledge, this is the first time that such tools
have been used to answer questions of this type.

While there has been extensive work on fair allocation of resources
over the years, there seems to be very little work that like us
tackles the fair allocation of multiple resources of distinct types.
A very recently suggested approach to this problem is dominant
resource fairness, where allocations are set so as to equalize each
user's maximal allocation of any resource \cite{ghodsi11}.
We discuss the similarities and differences between this scheme and
ours in Section \ref{sect:drf}.

\section{Prior Work}
\label{sect:prev}

To put our contributions in context, we first review prior work in
fair allocation of resources in systems.
The issue of resource allocation has been studied for many years, but
mostly from different perspectives than the one we use.

The requirement for control over the allocation of resources given to
different users or groups of users has been addressed in several contexts.
It is usually called \emph{fair-share scheduling} in the literature, where
``fair'' is understood as according to each user's entitlement,
rather than as equitable.
Early implementations were based on accounting, and simply gave
priority to users who had not yet received their due share at the
expense of those that had exceeded their share \cite{henry84,kay88}.
In Unix systems,  one approach that has been suggested
\cite{epema98,hellerstein93} is to manipulate each
process's ``nice'' value to achieve the desired effect.%
\footnote{``Nice'' is a user-controlled input to the system's priority
calculations. It is so called because normal users may only
\emph{reduce} their priority, and be nice to others.}
Simpler and more direct approaches include lottery scheduling
\cite{waldspurger94} or using an economic model \cite{stoica95}, where
each process's priority (and hence relative share of the resource) is
expressed by its share of lottery tickets or capital.

Another popular approach is based on \emph{virtual time}
\cite{duda99,nieh01}.
The idea is that time is simply counted at a different rate for
different processes, based on their relative allocations.
In particular, scheduling decisions may be based on the difference
between the resources a process has actually received and what it
would have received if the ideal processor sharing discipline had been
used \cite{bennun10,chandra00,etsion06}.
This difference has also been proposed as a way to measure
(un)fairness that combines job seniority considerations with resource
requirements considerations \cite{aviitzhak07,raz04}.

In networking research, control over relative allocations is achieved
using \emph{leaky bucket} or \emph{token bucket} metering approaches.
This is combined with fair queueing, in which requests from different
users are placed in distinct queues, which are served according to how
much bandwidth they should receive \cite{demers89,nagle87}.
The most common approach to fairness is max-min fairness, where the
goal is to maximize the minimal allocation to any user \cite{radunovic07}.

Focusing on virtual machine monitors, Xen uses a credit scheduler
essentially based on virtual time, where credits correspond to
milliseconds and domains that have extra credit are preferred over
those that have exhausted their credit \cite{ongaro08}.
Note, however, that domains that have gone over their credit limit may
still run, as in borrowed virtual time \cite{duda99}.
VMware ESX server uses weighted fair queueing or lottery scheduling
\cite{demers89,waldspurger94}.
The Virtuoso system uses a scheduler called VSched that treats virtual
machines as real-time tasks that require a certain slice of CPU time
per each period of real time \cite{lin05,lin06}.
Controlling the slices and periods allows for adequate performance even
when mixing interactive and batch jobs.

The main drawback of the approaches mentioned above is that they
focus on one resource --- the CPU, or in a networking context, the
bandwidth of a link.
The effect of CPU scheduling on I/O is discussed by Ongaro et
al.\ \cite{ongaro08} and Govindan et al.\ \cite{govindan07}.
For example, they suggested that VMs that do I/O could be
temporarily given a higher priority so as not to cause delays and
latency problems.
However, the interaction of such prioritization with allocations was
not considered.
Similarly, there has been interesting work on scheduling bottleneck
devices other than the CPU \cite{bansal01,harcholb03,schroeder06}, but
this was done to optimize performance of the said device, and not to
enforce a desired allocation.

Few works have considered dealing with \emph{multiple} resource constraints.
Diao et al.\ \cite{diao02} suggested an approach of
controlling applications so that they adjust their usage, rather than
to enforce an allocation.
Fairness in the allocation of multiple resources was addressed by
Sabrina et al.\ \cite{sabrina07} in the context of packet scheduling,
where the resources were network bandwidth and the CPU resources
needed to process packets.
The approach taken was to consider the processing and transmission
times together when using a weighted fair queueing framework.
The interaction between scheduling and multiple resources was
discussed by Amir et al.\ \cite{amir00}.
However, the context is completely different as they consider targets
for migration in the interest of load balancing.
Interestingly, the end result is similar to our approach, as they try
to avoid machines where any one of the resources will end up being
highly utilized and in danger of running out (and becoming a
bottleneck).
Control over multiple resources was also considered at the
microarchitectural level by Bitirgen et al.\ \cite{bitirgen08}, but
with a goal of achieving performance goals rather than predefined
allocations.

Our work extends a recent suggestion to focus on bottleneck resources
\cite{bennun10,etsion09}.
Specifically, the suggestion was to identify at each stage which device
is the system bottleneck (that is, the device whose usage
is closest to 100\% utilization) and then enforce the desired allocation
on this device. 
For example, if the disk is the bottleneck, one can promote or delay
requests from different users so as to achieve the desired
relative allocation of bandwidth among them.
This, in turn, induces corresponding usage patterns on other devices
including the CPU.
But if the disk is \emph{the} bottleneck, the other devices will be
less than 100\% utilized, and therefore scheduling them is less
important.
However, this suggestion does not deal with what to do if there are
in fact multiple bottlenecks, which, as we observed, can easily happen.
Our work extends the definition to cover the case of multiple
bottlenecks.

In networking, allocations to flows traversing multiple links are also
typically viewed as using multiple resources, where again the
constraints stem from links that become saturated (and hence a
bottleneck).
In this context min-max fairness can be characterized based on a
geometrical representation that is very similar to ours
\cite{radunovic07}.
However, the requirements from all the resources (links) are equal,
making the search for a solution easier.
Specifically, it is often possible to move in a straight line from the
origin to the boundary, in a direction based on the desired relative
allocations, rather than using a more complicated trajectory as we do
in Section \ref{sect:proof}.

To the best of our knowledge, the only other work to suggest and
analyze a fair-share allocation policy that handles diverse
requirements for multiple resources is the recently proposed dominant
resource fairness \cite{ghodsi11}.
This does not explicitly consider bottlenecks, but rather focuses on
each user's maximal usage of any single resource.
We describe this in more detail and compare it with our definition in
Section \ref{sect:drf}.

\section{Sharing Multiple Resources}
\label{sect:bottleneck}

Fair sharing of resources has been one of the objectives of
scheduling for many years, and has received renewed interest in the
contexts of virtualization and cloud computing.
But what exactly is ``fair sharing''?
Consider a setting with $N$ users and $m$ resources (CPU, network
bandwidth, disk usage, and so on).
We assume that each user $i$ is entitled to a fixed percentage $e_i$ of
the full capacity, and hence of each resource, where
$e_1 + \cdots + e_N = 1$.
Each user $i$ requests a fraction $r_{ij}$ of resource $j$.
If $r_{ij} < e_i$ for all $j$---that is, if $i$ requests less than his
entitlement on each resource---then any reasonable notion of fair
sharing should grant user $i$ all that he requests on each resource.

But what if user $i$ requests more than his entitlement on some
resource $j$?
In this case, if $r_{1j} + \cdots + r_{Nj} \le 1$ for each resource $j$, so
that no resource is a bottleneck, then any efficient notion of fair
sharing should give each user all that he requests.
Even if a user requests more than he is entitled to of some resource,
as long as no resource is a bottleneck, there is no problem.
Clearly the problem arises only when $r_{1j} + \cdots + r_{Nj} > 1$ for
some resource $j$.
If there is only one bottleneck resource, again it seems easy enough
to cut back those users who are requesting more than their
entitlement \cite{etsion09}.
But what if there are several bottlenecks?
What should ``fair allocation'' mean in this case?

The problem is compounded by the fact that different users have
different requirements $r_{ij}$ for the different resources.
For example, in an operating system setting, if a certain process is
entitled to 50\% of the 
resources, but this is an I/O-bound process that hardly uses the CPU,
the scheduler cannot force it to use more and fill its allocation.
Moreover, reserving 50\% of the CPU for this process will likely just
waste most of this capacity.
However, if we allocate the unused capacity to another process, which
also turns out to be I/O-bound, we may end up hurting the performance
of the original process.
We therefore need to find a set of allocations that allow us to
exploit complementary usage profiles to achieve high
utilization, but at the same time respect the different entitlements.
By respecting the entitlements, the allocations can be claimed to be
fair.
In particular, we define fairness by invoking the user's point of
view of the entitlements:
\begin{frameandlabel}{Fairness Definition}
  An allocation of multiple resources is fair if users have no
  justification to complain that they got less than they deserve.
\end{frameandlabel}

A key contribution of this paper is to define the properties of an
allocation that satisfies this definition, i.e.\ one where any
complaints would be unjustified.
We then go on to prove that such an allocation is in fact achievable,
for any combination of requirements and entitlements.
The discussion above already illustrates the core of our approach:
a focus on \emph{bottleneck} resources.
This approach is in line with basic results in performance evaluation,
as it is well known that the bottleneck device constrains system
performance (this is, after all, the definition of a bottleneck)
\cite{lazowska:book}.
An important manifestation of this result is that, in a queueing
network, most of the clients will always be concentrated in the queue
of the bottleneck device.
This implies that scheduling the bottleneck device is the only
important activity, and moreover, that judicious scheduling can be
used to control relative resource allocations.

Precisely this reasoning led to the recent suggestion that
proportional resource allocation be exercised on the bottleneck
device at each instant, rather than on the same device (e.g.\ the CPU)
at all times \cite{bennun10,etsion09}.
Focusing on the bottleneck in this way avoids trying to control
allocations based on an irrelevant tuning knob, and provides the most
reasonable interpretation of enforcing resource allocations in a
multi-resource environment.

But what happens if there are two or more bottlenecks?
In order to derive the allocations, we first need to define a model
of how resources are used.
Given the definitions of entitlements and requirements above,
our task is to figure out how much to cut each user back.
We assume that we cut each user back by the same factor $x_i$ on each
resource.
This is in fact our main assumption:
\begin{frameandlabel}{Proportional Resource Usage Assumption}
  Users use diverse resources in well-defined proportions.
  Thus cutting back on one resource by a certain factor will lead to
  reduced usage of other resources by the same factor.
\end{frameandlabel}
This assumption reflects a model where each user is engaged in a
specific type of activity with a well-defined resource usage profile.
For example, a user may be serving requests from clients over the
Internet.
Each request requires a certain amount of computation, a certain
amount of network activity, and a certain amount of disk activity.
If the rate of requests grows, all of these grow by the same factor.
\begin{figure}\centering
 \includegraphics[width=.48\textwidth]{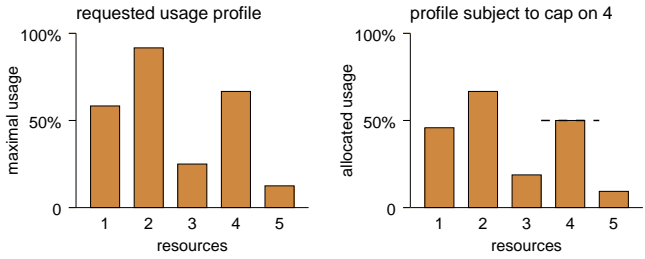}
  \caption{\label{fig:prof}\sl
    Illustration of a resource request profile, and how it is cut back
    when usage of resource 4 is limited to 50\%.}
\end{figure}
But if one resource is constrained, limiting the rate of serving
requests, this induces a similar reduction in the usage of all other
resources (see demonstration in Fig.\ \ref{fig:prof}).
This is essentially the ``knee model'' of Etsion et
al.\ \cite{etsion06}, where I/O activity is shown to be linearly
proportional to CPU allocation up to some maximal usage level.
It also corresponds to the task model of Ghodsi et al.\
\cite{ghodsi11} when all tasks that a user wants to execute have identical
resource requirements (which is indeed the specific model they use in
their proofs).
Note, however, that this is indeed a limiting assumption.
Specifically, it excludes usage patterns where one resource is used to
compensate for lack of another resource, as happens, for example, in
paging, or when using compression to reduce bandwidth.

All the above leads to the following problem definition.
We want to find $x_1, \ldots, x_N$, such that $0 \le x_i \le 1$, 
where for each user $i$, $x_i$ is the fraction of that user's
request which will be granted.
Furthermore, we require that
\begin{equation}\label{eq1}
  \mbox{for each resource } j: ~ x_1 r_{1j} + \cdots + x_N r_{Nj} \le 1.
\end{equation}
This means that the total usage of each resource is limited by the
resource capacity.
Those resources for which equality holds are the \emph{bottleneck}
resources.

Among all the allocations $x_1, \ldots, x_N$ that satisfy Eq.\ \eqref{eq1},
which should qualify as ``fair''?
This is where we define the ``no justified complaints'' condition:
\begin{frameandlabel}{No Justified Complaints Condition}
  A user cannot justify complaining about his allocation if either
  he gets all he asked for, or else he gets his entitlement, and giving
  him more would come at the expense of other users who have their own
  entitlements.
\end{frameandlabel}
Using the notation above, the ``no justified complaints'' condition can be
formally expressed as:
\begin{equation}\label{eq2}
  \begin{array}{l}
  \mbox{for all users } i: x_i = 1 \mbox{ or } \\ ~~~~
  \mbox{there exists a  bottleneck resource $j^*$ such that }
  x_i r_{ij^*} \ge e_i.
  \end{array}
\end{equation}
Specifically, user $i$ cannot complain if there exists some
bottleneck resource $j^*$ where he gets at least what he is entitled to.
Since $j^*$ is already utilized to its full capacity, giving him more,
that is, increasing $x_i$, would necessarily come at the expense of
other users, who have the right to their own entitlements.
Therefore, increasing $i$'s allocation at their expense would be
unfair.

Note that it may happen that a user receives less than his
entitlement on \emph{other} resources, including other bottleneck
resources, where the entitlement would seem to indicate that a larger
allocation is mandated.
This is where the proportional resource usage assumption comes in.
Recall that the factor $x_i$ is common to all resources.
Thus giving a user a higher allocation on any resource implies that
his allocation must grow on \emph{all} resources.
The original bottleneck resource $j^*$ thus constrains all
allocations, even on other bottleneck resources or resources that are
not themselves contended.

Being based on bottlenecks, it is easy to see that allocations that
satisfy Eqs.\ \eqref{eq1} and \eqref{eq2} are Pareto optimal 
(but, of course, not every Pareto-optimal solution satisfies our
fairness criterion).
Showing that such a fair allocation exists turns out to be
surprisingly nontrivial.
The obvious greedy approach does not seem to work. 
Given a collection of requests and entitlements, suppose that we try
to satisfy the users one at a time, so that, after the $k$th step, we
have an allocation $(x_1, \ldots, x_N)$ satisfying Eq.\ \eqref{eq1} such
that the first $k$ users have no complaints (that is, Eq.\ \eqref{eq2} 
holds for users $1, \ldots, k$).
To see why doing this does not seem helpful, suppose that there are
three resources and three users.
User 1 requests $(\frac 1 2, \frac 1 2, \frac 2 3)$
(i.e., $r_{11} = \frac 1 2$, $r_{12} = \frac 1 2$, and $r_{13} = \frac 2 3$)
and is entitled to $0.5$ of the resources (i.e., $e_1 = \frac 1 2$).
User 2 requests $(\frac 1 2, \frac 5 8, \frac 1 2)$ and $e_2 = \frac 3 8$,
and User 3 requests $(1,1, \frac 1 3)$ with $e_3 = \frac 1 8$. 
We start by giving user 1 everything he asks for, and users 2 and 3
nothing; that is, we consider the allocation $x = (1,0,0)$.
Clearly at this point User 1 has no complaints.
Next we try to satisfy User 2.
If we do not cut back User 1, then we must have $x_2 \le \frac 2 3$, since
resource 3 then becomes a bottleneck.
But with the allocation $x = (1, \frac 2 3, 0)$ User 2 has a justified
complaint: the only resource on which he gets at least his entitlement
is resource 2, but resource 2 is not a bottleneck with this allocation.
So User 2 feels that he is entitled to a bigger share of the
resources. 
There are various ways to solve this problem.
For example, we could consider the allocation $(\frac 3 4,1,0)$.
It is easy to check that neither User 1 nor User 2 has a justified
complaint with this allocation.
But now we need to add User 3 to the mix.
To do so, we have to cut back either User 1 or User 2, or both.
It follows from our main theorem that this can be done in a way that
none of the users has a justified complaint.
But the naive greedy construction does not work; at each stage, we
seem to have to completely redo the previous assignment.
Although a more clever greedy approach might work, we suspect not; 
a more global approach seems necessary.
We will describe such an approach in Section \ref{sect:exist}.

\section{Properties of Allocations with No Justified Complaints}
\label{sect:drf}

In their analysis of dominant resource fairness, Ghodsi et
al. \cite{ghodsi11} show that it possesses four desirable attributes,
under the assumption that all tasks that a user wants to execute have
identical resource requirements (in which case their model reduces to ours).
We now show that our definition possesses them as well.
This answers Ghodsi et al.'s question of whether there are other fair
allocation schemes with these properties.

The first attribute is providing an \emph{incentive for sharing}: the
allocation given to each user should be better than just giving him
his entitlement of each resource
(actually they defined this requirement only in the case that
all users are viewed as having equal entitlements, in which case this
amounts to giving each user $\frac 1 n$th of each resource).
Suppose that if user gets a fraction $e_i$ of each resource, he can
perform a fraction $x$ of his requests, where $xr_{ij} \le e_i$.  In an
allocation that is far in our sense, user $i$ gets to perform a fraction
$y$ of his reuqests, where $yr_{ij} = e_i$ for some resource $j$.  Thus,
we must have $x \le y$, which means that player $i$ is at least as well
off participating in the scheme as he would be if he got his entitlement
on each resource.

The second attribute is being \emph{strategyproof}.
This means that users won't benefit from lying about their resource
needs.
Asking for less than the real requirements obviously just caps the
user's potential allocation at lower levels.
Asking for more with the same profile (that is, same relative usage of
different resources) does not have an effect, except that the user
might be allocated more than he can use.
While this may lead to waste, it does not provide any benefit to the
user.
Modifying the profile will either give the user extra capacity he
can't use on some resource, or worse, reduce the effective allocation
because some unneeded resource was inflated and tricked the system
into thinking it has satisfied the user's entitlement.
Thus lying cannot lead to benefits, but can in fact cause harm to a
user's allocation.

The third attribute is that the produced allocation be \emph{envy
free}: no user should prefer another user's allocation.
This follows from being strategy proof; otherwise a user could lie about
his requirements so as to mimic those of the other user. 

The fourth and final attribute is \emph{Pareto efficiency}.
This means that increasing the allocation to one user must come
at the expense of another.
As noted above, this follows from doing allocations based on bottlenecks.

We now turn to comparing our definition of fairness with dominant
resource fairness.
While similar in spirit, the two definitions are actually quite
different in their philosophy.
At a very basic level, the notion of fairness depends on perception of
utility.
In the context of allocating resources on computer systems, the
utility is typically unknown.
Consequently the notion of fairness is ill-defined.

To better understand the difference between utility and allocation, we
recount an example used by Yaari and Bar-Hillel \cite{yaari84}.
Jones and Smith are to share a certain number of grapefruit and
avocados to obtain certain vitamins they need.
They have different physiological abilities to extract these vitamins
from the different fruit.
The overwhelming majority of those polled agreed that the most fair
division is one that gives them equal shares of extracted vitamins,
despite being quite far from being equal shares of actual fruit.
But such considerations would be impossible if you do not know their
specific ability to extract vitamins, and that they actually only eat
fruit for their vitamins.

When allocating resources to virtual machines or users of a cloud
system, we do not know the real utility of these resources for the
users.
We are therefore forced to just count the amount of resources being
allocated.
The difference between definitions of fairness is in how this counting
is done.
In asset fairness, the fractions of all resources used are summed up.
Thus if a user gets 20\% of the CPU, 7\% of the disk bandwidth, and
37\% of the network bandwidth, he is considered as having received
64/300 of the total resources in the system.
In order to be fair, other users should also get similar total
fractions.
In dominant resource fairness, only the largest fraction is
considered.
Thus, in the example above, the user's dominant resource is the network,
and he is considered to have received resources at a level of 37/100.
To be fair, other users should receive similar levels of their
respective dominant resources. 
In our definition of fairness, we do not focus on the dominant resource
of each user, but rather take a system-wide view based on bottlenecks.
Thus, if the CPU happens to be the only bottleneck, we say that
this user received resources at a level of 20/100.
The fact that he received more of another resource, namely the
network, is immaterial, because there is no contention for the network.
A user is welcome to use as much of any resource for
which there is no contention as he likes. 

\begin{figure}\centering
 \includegraphics[width=.4\textwidth]{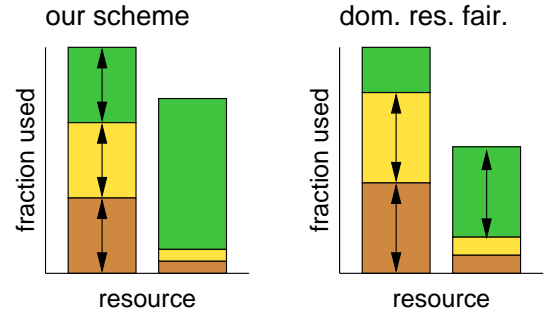}
  \caption{\label{fig:div1}\sl
    Example of the effect of imposing equal shares of a bottleneck
    resource, compared with dominant resource fairness.}
\end{figure}

Interestingly, Ghodsi et al.\ prove that under dominant resource
fairness each user will be constrained by some resource that is a
bottleneck \cite{ghodsi11}.
However, their fairness criterion does not depend on this bottleneck,
while ours does.
The following example may help to illustrate the differences
(Fig.\ \ref{fig:div1}).
Consider a scenario with three users and two resources.
The requirements of the users are $r_1 = (1, 0.2)$,
$r_2 = (1, 0.2)$, and $r_3 = (0.4, 0.8)$.
The entitlements are $e_1 = e_2 = e_3 = \frac 1 3$.
Obviously resource 1 is a bottleneck, so the allocations with our
definition of fairness will be $a_1 = (\frac{1}{3}, \frac{2}{30})$,
$a_2 = (\frac{1}{3}, \frac{2}{30})$,
and $a_3 = (\frac{1}{3}, \frac{2}{3})$.
This is fair on the bottleneck resource, and each user receives his
entitlement.
Dominant resource fairness, in contrast, leads to the following
allocation: $a_1 = (0.4 , 0.08)$, $a_2 = (0.4 , 0.08)$, and
$a_3 = (0.2 , 0.4)$.
User 3's usage of resource 2 is counted, despite the fact that there is
no contention for resource 2; this leads to a reduced allocation of
resource 1.
There seems to be no criteria by which to say that one allocation is
fairer than the other.
It may well be that user 3 derives much benefit from using resource 2,
and therefore cutting him back on resource 1 is perfectly justified.
But given that we do not \emph{know} that this is the case, we suggest
that it is safer to focus on the bottleneck resources.

In fact, Ghodsi et al.\ do mention bottleneck fairness in their
description of dominant resource fairness, but only as a secondary
criterion.
They define bottleneck fairness only when all users have the same
dominant resource, essentially reducing the scope to the single
bottleneck case.
Our work is the first to extend this with a meaningful definition of
fairness  for multiple bottlenecks, and when the dominant resources
are different.

We now turn to a few more observations of the relationship between our
definition and dominant resource fairness.
First, we observe that if all users have the same dominant resource,
dominant resource fairness and our definition are equivalent.
This follows since the common dominant resource  is the only bottleneck.

Another interesting question is one of utilization.
In the example given above, our definition of fairness led to higher
overall utilization than dominant resource fairness.
It this guaranteed to always be the case?
The answer is no, as the following counter-example demonstrates.
Assume two users and four resources, with requirement vectors of
$r_1 = (\frac 1 2, 0, 0, 1)$ and $r_2 = (1, 1, 1, 0)$ and
equal entitlements $e_1 = e_2 = \frac 1 2$.
With our no justified complaints definition, the first and last
resources are the bottlenecks, and the allocations are
$a_1 = (\frac 1 2, 0, 0, 1)$ and
$a_2 = (\frac 1 2, \frac 1 2, \frac 1 2, 0)$.
If there were many more ``middle'' resources, the average utilization
would tend to $\frac 1 2$.
With the dominant resource fairness scheme, the allocations are
$a_1 = (\frac 1 3, 0, 0, \frac 2 3)$ and
$a_2 = (\frac 2 3, \frac 2 3, \frac 2 3, 0)$.
In this case, the average utilization tends to $\frac 2 3$.

Another important difference between the two definitions is that
dominant resource fairness allocations can be found using an
incremental algorithm \cite{ghodsi11}.
Finding allocations based on the no justified complaints idea is
harder, because we do not know in advance which resources will be the
bottlenecks.
Nevertheless, the proof presented in the next section shows that such
an allocation always exists.
Moreover, the trajectory argument used is actually somewhat similar to
how allocations are constructed for dominant resource fairness.

\section{Existence of a Fair Allocation}
\label{sect:exist}

In this section, we prove that an allocation satisfying \eqref{eq1} and
\eqref{eq2} always exists.
Note, however, that there is an additional requirement that $x_i \le 1$,
and that \eqref{eq2} makes a distinction between the case $x_i < 1$
(need to exceed entitlement on some bottleneck resource) and the case
$x_i = 1$ (get all you want).
We can treat these two cases uniformly by defining $N$ dummy
resources that are each requested by only one user.
Using $m'$ to denote the number of \emph{real} resources,
we define the requirements on the dummy resources to be
$r_{i,m'+i} = 1$ for $i = 1,\ldots, N$, and
$r_{i,j} = 0$ for $i = 1,\ldots, N$, $j = m'+1, \ldots, m'+N$, and\
$j \ne m'+i$.
These dummy resources can only become a ``bottleneck'' if their
corresponding $x_i = 1$, meaning that the user gets all he requested.
In the following, $m$ will denote the full set of resources including
the dummy ones.

With this addition, we want to prove the following:
\begin{theorem}\label{thm:main}
  Given
\vspace*{-5mm}
  \begin{itemize}\itemsep 0pt
  \item entitlements $e_1, \ldots, e_N$ such that
    $e_1 + \cdots + e_N = 1$ and $e_i \ge 0$ for $i = 1, \ldots, N$,
    and
  \item resource requirements $r_{ij}$,
    $i = 1,\ldots, N$, $j = 1, \ldots, m$ such that 
    $r_{1j} + \cdots + r_{Nj} \ge 1$ for $j = 1, \ldots, m$
    and $0 \le r_{ij} \le 1$ for $i = 1,\ldots, N$ and $j=1,\ldots,m$, 
  \end{itemize}
\vspace*{-5mm}
there exists an allocation $x_1, \ldots, x_N$,
where $0 \le x_i \le 1$ for $i = 1, \ldots, N$,
that satisfies the two conditions
\vspace*{-5mm}
  \begin{enumerate}[(1)]\itemsep 0pt
  \item for all resources $j$, $j = 1, \ldots, m$, ~
    $x_1 r_{1j} + \cdots + x_N r_{Nj} \le 1$;
  \item for all users $i$, $i = 1, \ldots, N$,
    there exists a resource $j^*$
    such that $x_i r_{ij^*} \ge e_i$ and 
    $x_1 r_{1j^*} + \cdots + x_N r_{Nj^*} = 1$.
  \end{enumerate}
\end{theorem}
As the mathematical derivation is somewhat involved, we first provide
an argument for the special case $N=2$ (two users); this 
enables us to draw the constructions used in 2D.
The full proof for all values of $N$ is given in Section \ref{sect:proof}.

\subsection{Simplifying Assumptions}

Before proving the theorem, 
we make three simplifying assumptions, all without loss of generality.
First, as reflected in the definition of the resource requirements, we
assume that, for each resource $j$, $r_{1,j} + \cdots + r_{Nj} \ge 1$.
If there is any additional resource $j^\diamond$ for which
this inequality does not hold, we can ignore resource $j^\diamond$,
and solve the problem for the remaining resources.
Whatever solution we come up with will also be a solution when we add
$j^\diamond$ back to the picture, because its usage will be at most
$r_{1,j^\diamond} + \cdots + r_{Nj^\diamond} < 1$.

Second, we assume that, for each user $i$,
there is at least one resource $j$ such that $r_{ij} \ge e_i$.
(This pertains to only real resources, not the dummy resources.)
If this is not the case, we give user $i$ everything he asked for,
remove his requests, renormalize the entitlements of the remaining
users so that they still sum to 1,
renormalize the remaining capacity of the different resources so that
it is still 1, and renormalize the remaining requests by the same factors.
For example, suppose that users 1, 2, and 3
are entitled to 0.5, 0.2, and 0.3 of capacity, respectively.
If User 1 never  asks for more than 0.5 of any resource, then we give
him what he asks for, and remove his requests from the picture. 
Note that this means that, for each resource $r$, the fraction of $r$ 
available is at least as much as the entitlement of each user.
We then multiply User 2 and User 3's entitlements by 2
($= 1 / (1-0.5)$), so that their entitlements still sum to 1.
After this normalization, they are entitled to 0.4 and 0.6 of what
remains after we have granted User 1's request.
Moreover, if User 1 requested, say, 0.4 of Resource 1, so that 60\%
of Resource 1 is still available, we multiply each of the remaining
user's requests by $\frac 5 3$ ($= 1 / 0.6$).
Again, if we solve the resulting problem, we will have solved our
original problem.   
This follows in general since, if User 1 is the one eliminated,
the entitlements of the remaining users
effectively grew by a factor of $1/(1-e_1)$,
while the requests and capacity of resource $j$ grew by
$1/(1-r_{1j})$.
Since $r_{1j} < e_1$ the entitlements grew by a larger factor, and
fulfilling them will also satisfy the original entitlements.

Finally, we assume that there are no \emph{dominated} inequalities,
where an inequality $x_1 r_{1j} + \cdots + x_N r_{Nj} \le 1$ is
dominated if any solution $(x_1,\ldots, x_N)$ to the remaining
inequalities is also a solution to this inequality.
Dominated inequalities can be efficiently found by standard linear
programming methods.
We can clearly remove dominated inequalities to get a system with no
dominated inequalities.
Depending on the order of removal, we may end up with different
systems.
However, a solution to any of the undominated
systems is also a solution to the original system.

We now prove that we can find a solution $x_1 \ldots x_N$ satisfying
the requirements of Theorem \ref{thm:main} under these simplifying
assumptions.
We stress that this is without loss of generality; as shown above, if we
can find a solution under the simplifying assumptions, we can also find
one without these assumptions.

\subsection{Proof Structure}

\begin{figure}\centering
 \includegraphics[width=.3\textwidth]{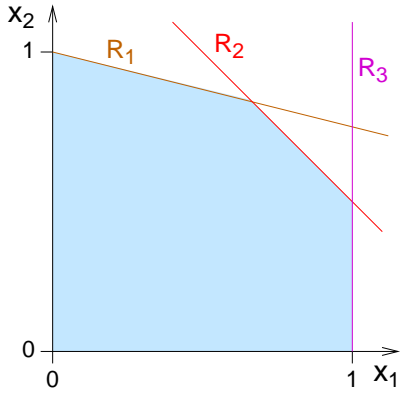}
  \caption{\label{fig:twoD}\sl
    Depiction of bounds on $x_i$ values due to capacity constraints of
    resources, for $N=2$ and $m=3$.}
\end{figure}

We first establish some notation.
By \eqref{eq1}, the set of legal allocations is a subset
$\calD$ of $\mathbb{R}^N$, where
\[
\begin{array}{l@{\,}l}
\calD = \{\: & (x_1,\dots,x_N):\,\, 0 \le x_i \le 1, \,\,\forall i
	\quad\text{ and }\quad \\
      & x_1 r_{1j} + \cdots + x_N r_{Nj}\le 1, \,\,\forall j \:\}. \\
\end{array}
\]
For $N=2$, this is a polygon in the first quadrant, as illustrated in
Fig.\ \ref{fig:twoD}.
In the figure, two users contend for three resources ($m=3$).
The request vectors are $r_1 = (\frac 1 4, \frac 2 3, 1)$
and $r_2 = (1, \frac 2 3, 0)$.
This leads to the bounds shown;
for example, the point $(\frac 2 3, \frac{11}{12})$ is impossible because
it would imply using
$\frac 2 3 \cdot \frac 1 4 + \frac{11}{12} \cdot 1 = \frac{13}{12}$ of
resource 1, i.e.\ more than its capacity.
In the general case, this region is a simplex in the positive orthant
(that is, the convex hull of $N$ affinely independent points, all in
$(\mathbb{R}^+)^{N}$).

For every vector $\x = (x_1,\dots,x_N)$ in $\calD$, the set of
bottleneck resources is
\[
J(\x) = \{j:\,\, 1\le j \le m, \quad x_1 r_{1j} + \cdots + x_N r_{Nj}=1\}.
\]
$J(\x)$ is empty for all $\x$ in the interior of the domain $\calD$,
implying that our solution will lie on the boundary of $\calD$.
Using this notation to re-write requirement \eqref{eq2}, our goal is
to prove that there exists an allocation $\x = (x_1,\dots,x_N)$, such
that
\begin{equation}\label{bakbuk}
  \begin{array}{l}
\mbox{for every~}~i=1,\dots,N~ \\ ~~~~~
\mbox{~there exists a~} j\in J(\x)
\mbox{~such that~} x_i r_{ij} \ge e_i.
  \end{array}
\end{equation}

The difficulty in finding $\x$ stems exactly from this condition.
In fact, if we knew what the bottleneck resources would be, the
problem could be solved efficiently using well-known machinery.
Specifically, fix an arbitrary subset $I \subseteq \{1,\ldots,m\}$, and
consider the following decision problem: Is there an $\x \in \calD$
for which $J(\x)=I$ such that condition~\eqref{bakbuk} holds?
It can easily be verified that this is asking whether a finite set of
linear equations and linear inequalities is consistent.
This task is subsumed by the Linear Programming problem, and can thus
be solved in polynomial time.

How can we overcome the difficulty involved in satisfying
condition~\eqref{bakbuk} without knowing in advance what the set
$J(\x)$ is?
We take a somewhat unconventional approach to this problem.
The set $\calD$ is a polytope, that is, a bounded convex subset of
$\mathbb{R}^N$ that is defined by a finite list of linear
inequalities.
We want to approximate $\calD$ by a subset $\cal{Q} \subseteq \calD$
that is convex and has a smooth boundary.
Intuitively, $\cal{Q}$ ``rounds off'' the corners of $\calD$ (see below
for further discussion).
Such a set $\calQ$ is defined by {\em infinitely many} linear
inequalities: For every hyperplane $H$ that is tangent to $\calQ$ we
write a linear inequality that states that $\x$ must reside ``below''
$H$.
It would seem that this only complicates matters, replacing the
finitely defined $\calD$ by $\calQ$.
However, the problematic condition~\eqref{bakbuk} takes on a much
nicer form when applied to $\calQ$, and becomes a very simple
relation involving the contact point of $H$ and $\calQ$, the normal
to $H$, and the vector $e$ (see Equation \eqref{eq:xne} below).
Moreover, using standard tools from the theory of ordinary
differential equations, we can find
a point on the boundary of $\calQ$ where this relation holds.

To find the solution, we do not consider a single smooth $\calQ$, but
rather a whole parametric family $\calQ_t$.
This family has the properties that (a) the sets $\calQ_t$ grow as the
parameter $t$ increases; (b) they are all contained in $\calD$; and
(c) as $t \rightarrow \infty$ the sets $\calQ_t$ converge to $\calD$.
For every $t > 0$, we find a point $\x^{(t)}$ on the boundary of
$\calQ_t$ such that $\x^{(t)}$ satisfies the analogue of
condition~\eqref{bakbuk}.
As $t \rightarrow \infty$ the points $\x^{(t)}$ tend to the boundary
of $\calD$.
We argue that there always exists a convergent subsequence of the
points $\x^{(t)}$, and show that the limit point of this subsequence
solves our original problem.
In the language of the description below, $\calQ_t$ is defined as
the set of those $\x \in \calD$ for which $f(\x) \le t$.

The procedure above hinges on our ability to define the appropriate
points $\x^{(t)}$ that satisfy the required condition.
This is based on considering the tangent to the surface of $\calQ_t$.
Note that the only essential difference between $\calD$ and $\calQ$ is
that the latter is defined by an infinite family of defining linear
inequalities, namely, one for each hyperplane $H$ that is tangent to
$\calQ$.
Keeping this perspective in mind, let us apply the original problem
definition to a point $\x \in \calQ$.
If $\x$ lies in the interior of $\calQ$,
then none of $\calQ$'s defining inequalities holds with equality.
Thus, as before, $J(\x)$ is empty for any $\x$ in the interior of the
domain $\calQ$.
We therefore consider $\x$ that lies on the boundary of $\calQ$.
In this case the set $J(\x)$ is a singleton, the only member of which
is the inequality corresponding to the hyperplane $H$ that is tangent
to $\calQ$ and touches it at the point $\x$.
The equation of the tangent hyperplane $H$ can be written as
$\sum \nu_i x_i = 1$, where the vector
$(\nu_1,\ldots,\nu_n)$ is normal to $H$.
Now condition~\eqref{bakbuk} becomes
\begin{equation}\label{pkak}
 \forall i~~~\nu_i x_i \ge e_i.
\end{equation}
When we sum over all $i$ this becomes
$\sum \nu_i x_i \ge \sum e_i = 1$.
But $\x$ lies on $H$, so that $\sum \nu_i x_i = 1$.
It follows that all inequalities in Eq.\ \eqref{pkak} hold with
equality.
But we also have, from the definition of the bottlenecks, that
$\sum r_{ij} x_i = 1$.
Thus, the normal is simply defined by the requirements vectors.
Moreover, we can use this as a condition on the gradients of the
surfaces of $\calQ_t$ for successive $t$'s, and follow a trajectory that
leads to a solution on the boundary of $\calD$.
This is then the desired constructive proof: it both shows that a
solution exists, and provides a mechanism for finding it.

\subsection{The Case \boldmath$N=2$}
\label{proofn2}

In this section, we give a complete proof of Theorem~\ref{thm:main} for
the case $N=2$ that is simpler than our general proof, and is perhaps
more intuitive.
This includes an explanation of the relationship between the normals to
the surfaces and the requirements vectors.
The argument for arbitrary $N$ is given in the next subsection. 

In the case $N=2$,
as noted above, the constraint \eqref{eq1} defines a 
region in the first quadrant whose boundary is a piecewise linear
curve that satisfies the constraints in \eqref{eq1} with $\le$
replaced by $=$.
Note that the slopes of the lines that define the
boundary are negative, and as $k$ increases from 0 to 1, the slopes of
the lines that intersect the vertical line $x=k$ get more and more
negative.
This follows from the fact that the interior is convex.

\begin{figure}\centering
 \includegraphics[width=.5\textwidth]{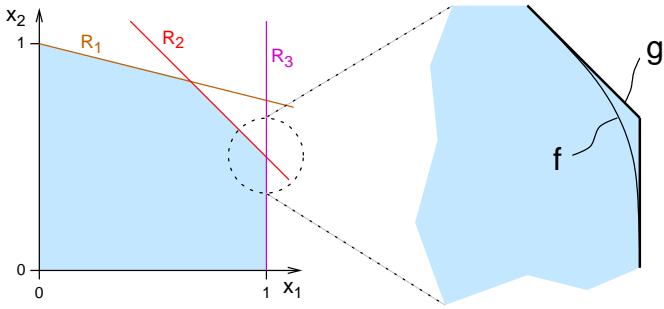}
  \caption{\label{fig:round}\sl
    Rounding off the boundary of $\calD$.}
\end{figure}

Let $g$ be the piecewise linear curve that defines the boundary.
We can approximate $g$ arbitrarily closely from below by a concave
twice-differentiable function $f$.
(The function $g$ is the boundary of the region called $\calD$ in the
previous section; the function $f$ is the boundary of the region $\calQ$.)
The concavity of the curve $f$ just means that $f'' < 0$.
As we said earlier,
$f$ ``rounds off''  the corners of $g$, as shown in
Fig.\ \ref{fig:round}.

As we said in the previous section, we want to find a point 
$(x_1^*,x_2^*)$ on the curve $f$ such that if
$\nu_1 x_1 + \nu_2 x_2 = 1$ is the tangent to the curve at
$(x_1^*,x_2^*)$, then $\nu_1 x_1^* = e_1$ and $\nu_2 x_2^* = e_2$.
We show below how to find such a point.  We now argue that finding such
a point for each $f$ approximating $g$ suffices to prove the theorem in
the case that $N=2$. 
First suppose that $(x_1^*,x_2^*)$ is
actually a point on one of the lines that define $g$ (as opposed to a
point on $f$ that arises from rounding off a corner of the curve $g$).
Suppose that the line is defined by resource $j$, so that it has the
form $r_{1j} x_1 + r_{2j} x_2 = 1$.
Obviously the tangent to $f$ at the point $(x_1^*,x_2^*)$ on the
line is just the line itself, so we have 
$\nu_1 = r_{1j}$ and $\nu_2 = r_{2j}$.
Thus, we will have found $x_1^*$ and $x_2^*$ such that
$r_{1j} x_1^* = e_1$ and $r_{2j} x_2^* = e_2$, which means that
\eqref{eq2} holds (and moreover, the same resource provides
justification for both users).

Next, suppose that $(x_1^*,x_2^*)$ is not on one of the original
lines, but we can find such a point for all functions $f$
approximating $g$.
Straightforward continuity arguments show that small changes to $f$
result in small changes to the point, so that as $f$ gets closer to 
$g$, we get a sequence of points that approach a point on $g$.  
Thus the limit is a point on $g$.  
We actually get even more.
What we really have for each function $f$
that approximates $g$ is two pairs $(x_1^*,x_2^*)$ and
$(\nu_1,\nu_2)$, where $(x_1^*,x_2^*)$ is a point on $f$,
$\nu_1 x_1 + \nu_2 x_2 = 1$ is the tangent to $f$ at $(x_1^*,x_2^*)$,
$\nu_1 x_1^* = e_1$, and $\nu_2 x_2^* = e_2$.
As $f$ approaches $g$, we will get a sequence of such pairs of points.
Let $(x^g_1,x^g_2)$ and $(\nu_1^g,\nu_2^g)$ be the limit of this sequence
of pairs of pairs.
It is clearly the case that $(x^g_1,x^g_2)$ is a point on $g$,
$\nu_1^g x_1^g = e_1$, $\nu_2^g x_2^g = e_2$, and 
$\nu_1^g x_1 + \nu_2^g x_2 = 1$.
Now if $(x^g_1,x^g_2)$ is in the interior of one of the lines
that make up the boundary of the region ---
let's assume it is the line associated with resource $j$ --- then, as
the argument above suggests, $\nu_1^g = r_{1j}$ and 
$\nu_2^g = r_{2j}$.
Thus, resource $j$ is a bottleneck, and provides a justification for
both users.

Now suppose that $(x^g_1,x^g_2)$ is at the intersection of two lines,
say, representing resources $j$ and $j'$.
Thus, $x^g_1 r_{1j} + x^g_2 r_{2j} = 1$ and
$x^g_1 r_{1j'} + x^g_2 r_{2j'} = 1$, so both resources are bottlenecks
at $(x^g_1,x^g_2)$.
Moreover, we still have $\nu_1^g x_1^g = e_1$ and $\nu_2^g x_2^g = e_2$.
Finally, it is clear that $\nu_i^g$ must be a convex combination of
$r_{ij}$ and $r_{ij'}$, for $i \in \{1,2\}$, since, for each
approximation $f$ to $g$, the tangent in the region that we have
``rounded off'' is a convex combination of the tangents of the lines
that make up $g$ that are being approximated.
It follows that each user $i \in \{1,2\}$ gets at least his
entitlement on one of resources $j$ or $j'$; that is, either 
$r_{ij} x_i^g \ge e_i$ or $r_{ij'} x_i^g \ge e_i$.
For if $r_{ij} x_i^g < e_i$ and $r_{ij'} x_i^g < e_i$,
then $\nu_i^g x_i^g < e_i$, and we have a contradiction.

The fact that we can find such a point $(x_1^*,x_2^*)$ on each $f$
follows from another easy continuity argument.
Consider the points on the function $f$ in the first quadrant.
Suppose that $f$ starts at the $Y$-axis at some
point $(0,y')$ and ends at the $X$-axis at some point $(x',0)$.  
Let the equation of the tangent of 
$f$ at the point $\x^\diamond = (x_1^\diamond,x_2^\diamond)$ be
$\nu^{\x^\diamond} \cdot \x = 1$.
Consider the term
$q(\x^\diamond) =
(\nu^{\x^\diamond}_1 x_1^\diamond) / (\nu^{\x^\diamond}_2 x_2^\diamond) =
- f'(x_1^\diamond) x_1^\diamond/x_2^\diamond$ 
as $\x^\diamond$ goes from $(0,y')$ to $(x',0)$.
As $\x^\diamond$ approaches $(0,y')$ from the right, 
$q(\x^\diamond)$ approaches 0;
as $\x^\diamond$ approaches $(x',0)$ from the left,
$q(\x^\diamond)$ approaches $\infty$.
Since $f'$ is continuous, $q$ varies continuously in the first
quadrant between $0$ and $\infty$.
Thus, at some point it must have value $e_1/e_2$.
If $q(\x^*) = e_1/e_2$, then we must have
$\nu^{\x^*}_1 x_1^* / \nu^{\x^*}_2 x_2^* = e_1 / e_2$.
Since we also have $\nu^{\x^*}_1 x_1^* + \nu^{\x^*}_2 x_2^* = 1$
and $e_1 + e_2 = 1$, it easily follows that we must have
$\nu^{\x^*}_1 x_1^* = e_1$ and $\nu^{\x^*}_2 x_2^*  = e_2$, as
desired.
This completes the proof in the case that $N=2$.

We can actually say more in the case that $N=2$.
Since $f$ is concave, $f'$ is decreasing, so $-f'$ is increasing.
It easily follows that $q$ is an increasing function.
Thus, there is a \emph{unique} point $(x_1^*,x_2^*)$ with the desired
properties.
It easily follows that, in the case of two users, the
solution  to \eqref{eq1} and \eqref{eq2} is unique.
Uniqueness has an important consequence.
In the problem definition, the bottleneck resources in \eqref{eq1} are
not known in advance.
In particular, it might seem that different solutions may lead to
different resources becoming bottlenecks.
Uniqueness guarantees that this is not the case, and that the set of
resources that will become bottlenecks is uniquely defined by the
problem parameters (that is, the entitlements and request profiles).
We remark that the uniqueness claim does not hold in general
for $N > 2$; see Section~\ref{sec:unique}.

\begin{figure}\centering
 \includegraphics[width=0.35\textwidth]{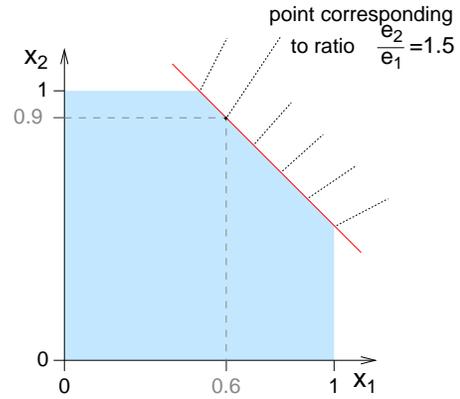}
  \caption[]{\label{fig:slope}\sl
  Simple example of a point $(x_1^*,x_2^*)$.
  Different points correspond to different ratios $\frac{e_2}{e_1}$,
  as indicated by the slopes of the line segments.}
\end{figure}

The example in Fig.\ \ref{fig:slope} may help in gaining an intuition
for the derivation above.
Consider a single limiting resource, where both
users request $r_1 = r_2 = \frac 2 3$ of its capacity.
The boundary line representing the capacity limit of the resource has
the equation $\frac 2 3 x_1 + \frac 2 3 x_2 = 1$.
Different points along this line correspond to different ratios of the
users' entitlements.
For example, if they have entitlements of $0.4$ and $0.6$, the point
$(0.6, 0.9)$ satisfies the equations $\frac 2 3 \cdot 0.6 = 0.4$ and
$\frac 2 3 \cdot 0.9 = 0.6$, and indeed using these values for $x_1$
and $x_2$ leads to sharing the resource in the desired proportions.
If the ratio of entitlements is such that $\frac{e_2}{e_1} > 2$,
then user 2 is not requesting his full entitlement, and is eliminated
from consideration.
He is given his full request, and user 1 gets the rest, which is more
than his entitlement (so they are both satisfied).
This is in fact an example of a solution based on a dummy resource.
The opposite happens if $\frac{e_2}{e_1} < \frac 1 2$.

\subsection{Proof of Theorem~\ref{thm:main}}
\label{sect:proof}

We now prove Theorem~\ref{thm:main} for arbitrary $N$.

\begin{construction}
To every allocation $\x$ in the interior of the domain $\calD$, we
assign a value
\begin{equation}\label{eq:f}
  f(\x) = -\sum_{j=1}^m \log\brk{1 - \sum_{k=1}^N x_k r_{kj}}.
\end{equation}
\end{construction}
\begin{remark}
The function $f$ is positive in
the interior of $\calD$, diverging to infinity as $\x$ tends to
the boundary of $\calD$.%
\footnote{Note that this function $f$ is not the curve $f$ of
  Section~\ref{proofn2}.}
\end{remark}
\begin{remark}
Clearly, there are other choices of $f$ that satisfy these desired
properties.
This choice seems like the simplest one for our purposes.
\end{remark}

\begin{figure}\centering
  \psfrag{infty}{$\infty$}
  \includegraphics[width=0.3\textwidth]{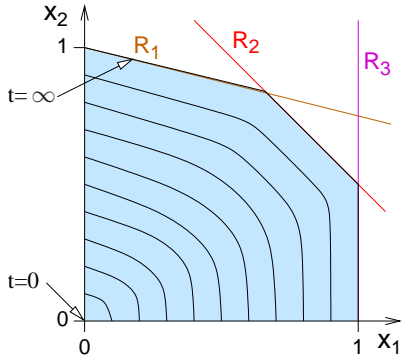}
  \caption[]{\label{fig:waves}\sl
    Illustration of level-sets of $f$ from $t=0$ to $t=\infty$.}
\end{figure}

\begin{definition}
To every number $t>0$, there corresponds a level set of $f$, namely,
\[
\Gamma_t = \{\x\in\calD:\,\, f(\x)=t\},
\]
\end{definition}
\begin{remark}
This is an $(N-1)$-dimensional hypersurface.
(Fig.\ \ref{fig:waves} illustrates this for $N=2$.)
\end{remark}

\begin{definition}
To every point $\x\in\calD$, there corresponds a unique unit vector
$\n(\x) = (\nu_1(\x),\dots,\nu_N(\x))$, normal to the level set of
$f$ at $\x$.
\end{definition}
The unit normal $\n(\x)$ is proportional to the gradient of $f$ at $\x$,
implying that 
\begin{equation}
\nu_i(\x) = \tilde{c}\,\pd{f}{x_i}(\x) =
  \tilde{c} \sum_{j=1}^m \frac{r_{ij}}{1 - \sum_{k=1}^N x_k r_{kj}},
\quad \forall i=1,\dots,N,
\label{eq:n}
\end{equation}
where the normalization constant $\tilde{c}$ is chosen so as to
guarantee that $\n$ is a unit vector, that is,
$\nu_1^2 + \dots + \nu_N^2 =1$.

\begin{construction}\label{construct:x(t)}
We now construct a vector-valued function
\[
\x(t) = (x_1(t),\dots,x_N(t)), \qquad t\ge 0,
\]
satisfying the following properties:\\[-5mm]
\begin{enumerate}\itemsep 0pt
\item $\x(t)$ lies on the level set $\Gamma_t$ for all $t\ge0$ 
  (and, in   particular, remains in $\calD$). 
\item For all $t>0$, there exists a $t$-dependent normalization factor
  $c(t)$, such that for every $i=1,\dots,N$,
  \begin{equation}\label{eq:xne}
    x_i(t)\, \nu_i(\x(t)) = \tilde{c}\,c(t) e_i.
  \end{equation}
\end{enumerate}
\end{construction}

\begin{remark}
Note that since $f(\x(0))=0$ it follows that $\x(0)=0$, that is, the
vector-valued function $\x(t)$ ``starts'' at the origin.
\end{remark}
\begin{remark}
substituting \eqref{eq:n} into \eqref{eq:xne} and summing over the
index $i$ determines $\tilde{c} \,c(t)$.
After simple algebraic manipulations, summing the expressions
\eqref{eq:xne} over $i$ gives us
\begin{equation}
\sum_{j=1}^m \frac{x_i(t) r_{ij} - (\sum_{k=1}^N x_k(t) r_{kj}) e_i}
                  {1 - \sum_{k=1}^N x_k(t) r_{kj}} = 0,
\; \forall i=1,\dots,N, \,\, \forall t>0 .
\label{eq:useful}
\end{equation}
\end{remark}

Intuitively, $\x(t)$ is a ``trajectory'' that takes us from the
origin $\x = 0$ to a point on the boundary of $\calD$ as $t$ grows
from $0$ to $\infty$.

The formal proof now follows from the following sequence of three
lemmas, proved below.
First, we show that a trajectory with the required properties exists
(Lemma~\ref{lemma:trajectory}). 
Given such a trajectory, we show that a subsequence of this trajectory
converges to a point on the boundary of $\calD$
(Lemma \ref{lemma:converge}).
Finally, this accumulation point is shown to be a solution to our
allocation problem (Lemma \ref{lemma:solve}).

It is convenient to delay the question of whether there indeed exists a
trajectory $\x(t)$ satisfying the required properties, and consider
convergence first.

\begin{lemma}\label{lemma:converge}
Let $0 < t_1 < t_2 < \cdots$ be a sequence tending to infinity.  
Let $x(t)$ be a vector-valued function as defined in Construction
\ref{construct:x(t)}.
Then, the sequence $x(t_i)$ has a subsequence that converges to an
allocation $\x^*$ on the boundary of $\calD$.
\end{lemma}
\proof
Consider what happens as $t\to\infty$.
Since $\x(t)\in\Gamma_t$, it follows that $\x(t)$ approaches the
boundary of $\calD$.
However, the function $\x(t)$ may not tend to a limit as $t\to\infty$.
Nevertheless, since $\calD$ is a compact domain, $\x(t)$ has
a convergent subsequence.
That is, there exists an allocation
$\x^* = (x_1^*,\dots,x_N^*)$ on the boundary of $\calD$
and a subsequence $t_{n_1}<t_{n_2}<\dots$
such that
\[
\lim_{k\to\infty} \x(t_{n_k}) = \x^*.
\]
\qed

The next lemma shows that this accumulation point is a solution to
the fair allocation problem.

\begin{lemma}\label{lemma:solve}
An allocation $\x^*$ as resulting from Lemma \ref{lemma:converge} is a
fair allocation according to our definition.
\end{lemma}
\proof
Since $\x^*$ is on the boundary of $\calD$, it has a non-empty 
set $J(\x^*)$ of bottleneck resources such that
\[
x^*_1 r_{1j} + \dots + x_N^* r_{Nj}=1
\qquad \forall j\in J(\x^*)\neq\emptyset.
\]
We then rewrite \eqref{eq:useful} by splitting the resources $j$ into
bottleneck resources and non-bottleneck resources, and setting $t=t_n$:
\begin{equation}\label{eq:4}
  \begin{array}{l}
\displaystyle
\sum_{j\not\in J(\x^*)} \frac{x_i(t_n) r_{ij} - (\sum_{k=1}^N x_k(t_n) r_{kj}) e_i}
			   {1 - \sum_{k=1}^N x_k(t_n) r_{kj}} \: + \\[6mm]
\displaystyle
\sum_{j\in J(\x^*) } \frac{x_i(t_n) r_{ij} - (\sum_{k=1}^N x_k(t_n) r_{kj}) e_i}
			  {1 - \sum_{k=1}^N x_k(t_n) r_{kj}} = 0.    
  \end{array}
\end{equation}
The two summations behave very differently as $n \to \infty$.
For a non-bottleneck resource $j$,  $\sum_{k=1}^N x_k^* r_{kj} < 1$, so
the summation over the non-bottleneck resources tends to a limit
obtained by letting $\x(t_n) \to \x^*$ term-by-term:
\begin{equation}\label{eq:5}
  \begin{array}{l}
\displaystyle
\lim_{n\to\infty} \sum_{j\not\in J(\x^*)}
\frac{x_i(t_n) r_{ij} - (\sum_{k=1}^N x_k(t_n) r_{kj}) e_i}
     {1 - \sum_{k=1}^N x_k(t_n) r_{kj}} \\[6mm]
\displaystyle
= \sum_{j\not\in J(\x^*)} \frac{x_i^* r_{ij} - (\sum_{k=1}^N x_k^* r_{kj}) e_i}
			   {1 - \sum_{k=1}^N x_k^* r_{kj}}.
  \end{array}
\end{equation}
For a bottleneck resource $j$, the denominator $1 - \sum_{k=1}^N x_k
r_{kj}$ tends to zero as $x\to x^*$, so the limit exists only if the
numerator vanishes as well.
But if it were the case that, for a given user $i$,
\[
x^*_i r_{ij} < e_i
\qquad \mbox{ for all } j\in J(\x^*),
\]
then
\[
\lim_{n\to\infty} \sum_{j\in J(\x^*) }
  \frac{x_i(t_n) r_{ij} - (\sum_{k=1}^N x_k(t_n) r_{kj}) e_i}
       {1 - \sum_{k=1}^N x_k(t_n) r_{kj}} = -\infty.
\]
This is a contradiction to the fact that, by \eqref{eq:4}, the limit
should be the negative of the right-hand side of \eqref{eq:5}.
Hence we conclude that $\x^*$ has the property that for all users $i$,
there exists a bottleneck resource $j$ such that $x^*_i r_{ij} \ge e_i$.
Thus, $\x^*$ is a fair allocation.
\qed

It remains to show that the trajectory $\x(t)$ is indeed well-defined
for all system parameters $e_i$ and $r_{ij}$.
This is handled by the following lemma.

\begin{lemma}\label{lemma:trajectory}
There exists a function $\x(t)$ with the properties specified in Construction
\ref{construct:x(t)}.
\end{lemma}
\proof
To prove this we show that we can find points satisfying property 1
that also satisfy property 2.
Since $\x(t) \in \Gamma_t$, we have $f(x(t)) = t$, that is,
\begin{equation}\label{eq:diff1}
-\sum_{j=1}^m \log\brk{1 - \sum_{k=1}^N x_k(t) r_{kj}} = t.
\end{equation}
By \eqref{eq:xne},
\begin{equation}\label{eq:diff2}
\sum_{j=1}^m \frac{x_i(t) r_{ij}}{1 - \sum_{k=1}^N x_k(t) r_{kj}} = c(t) e_i,
\qquad \forall i=1,\dots,N.
\end{equation}
Differentiating both equations with respect to $t$, we obtain a linear
system of equations for the derivative $d\x/dt$.
Differentiating \eqref{eq:diff1}, we get
\[
\frac{ \sum_{k=1}^N \deriv{x_k}{t} r_{kj}}{1 - \sum_{k=1}^N x_k(t) r_{kj}} = 1.
\]
Differentiating \eqref{eq:diff2}, we get
\begin{equation}\label{eq:linear}
\sum_{j=1}^m \frac{\deriv{x_i}{t} r_{ij}}{1 - \sum_{k=1}^N x_k r_{kj}} +
\sum_{j=1}^m \frac{x_i r_{ij} \sum_{k=1}^N \deriv{x_k}{t} r_{kj}}{(1 - \sum_{k=1}^N x_k r_{kj})^2}
 = \deriv{c}{t} e_i.
\end{equation}
Observe that, without loss of generality, we can set $dc/dt=1$,
compute the resulting 
vector of derivatives $d\x/dt$, and then multiply it by a constant for
the normalization condition to hold.
Thus, it remains only to show that \eqref{eq:linear} has a unique
solution when $dc/dt=1$.
To do so, we define an $\x$-dependent matrix with entries
\[
b_{ij} = \frac{r_{ij}}{1 - \sum_{k=1}^N x_k r_{kj}},
\qquad i=1,\dots,N, \qquad j=1,\dots,m\;.
\]
These entries are non-negative for $\x\in\calD$.
We now rewrite \eqref{eq:linear} in a more compact form,
\[
\sum_{k=1}^m \deriv{x_k}{t}  \brk{ \sum_{j=1}^m b_{ij} \delta_{ik} +
  \sum_{j=1}^N  x_i b_{ij} b_{kj}}  = e_i.
\]
The term inside the brackets is the $(k,i)$ entry of a symmetric
positive-definite $N\times N$ matrix, which immediately implies that
there exists a unique solution $d\x/dt$.
Moreover, since the dependence of $d\x/dt$ on $\x$ is continuous, the
existence and uniqueness of $\x(t)$ follows from the Fundamental
Theorem of Ordinary Differential Equations \cite{coddington:ode}.
(More precisely, the
fundamental theorem of ODEs guarantees only the existence and
uniqueness of a solution for some small $t$; global existence follows
from the boundedness of the domain $\calD$.)
\qed

This completes the proof of Theorem~\ref{thm:main}.

We note that our proof that a fair allocation exists is
almost constructive.
The trajectories $\x(t)$ can easily be computed numerically using
standard ODE integrators (for example, Matlab's \verb=ode45= function).
If $\x(t)$ is found to tend to a limit for large $t$, then this limit
is a fair allocation.
The only reservation is that numerical integration only provides
approximate solutions (however, with a controllable error), and can
only be carried out over a finite $t$ interval.

\subsection{Uniqueness of the Solution}
\label{sec:unique}

As we mentioned in Section~\ref{proofn2},
unlike the case $N=2$, in the general case the solution is not unique.
This is easily seen from the following counterexample.
Assume $N=3$ and $m=2$, with $r_1 = (1, 1)$, $r_2 = (0, 1)$,
$r_3 = (1, 0)$, and $e = (0.5, 0.3, 0.2)$.
This has the the family of solutions $x = (z, 1-z, 1-z)$ for $z$ that
satisfies $0.5 \le z \le 0.7$,
where in all these solutions both resources are bottlenecks.
Note that this does not contradict the fact that our solution
method finds a unique trajectory.
This trajectory corresponds to the choice of the function $f$ in
\eqref{eq:f}.
Other choices, e.g.\ by adding different weighting factors to each
term in the sum, could lead to other trajectories and other
solutions.

There also exist cases where different solutions depend on different
bottlenecks.
Consider the following example, with four users and four resources
($N = m = 4$).
Assume all users have the same entitlements, that is $e_i = 0.25$ for
$i = 1, \ldots, 4$.
Arrange the users and resources in a circle, and make each user
request the full capacity of its resource and those of its neighbors.
Thus the requirements matrix becomes
\[
r = \left(
\begin{array}{cccc}
  1 & 1 & 0 & 1 \\
  1 & 1 & 1 & 0 \\
  0 & 1 & 1 & 1 \\
  1 & 0 & 1 & 1 \\
\end{array}
\right)
\]
This instance is completely symmetric, and the obvious solution is a
symmetric allocation where $x_i = \frac 1 3$ for $i = 1, \ldots, 4$.
In this solution, all 4 resources are bottlenecks, and all users get
more than their entitlements on all the resources they use.
But there are 6 additional solutions.
Pick any two users $i$ and $j$, and set $x_i = x_j = 0.25$.
Let $k$ and $l$ be the other two users, and set $x_k = x_l = 0.375$.
Now two resources are bottlenecks ($0.25 + 0.375 + 0.375 = 1$) but the
other two are not ($0.25 + 0.25 + 0.375 = 0.875$).
Which resources become bottlenecks depends on the choice of $k$ and $l$.
If they are adjacent, then resources $k$ and $l$ are the bottlenecks.
If they are opposite each other, then resources $i$ and $j$ are the
bottlenecks.
In any case, every user gets his entitlement on at least one
bottleneck resource.
This demonstrates that the set of bottleneck resources is not unique.

The finding that there may be multiple solutions opens the issue of
selecting among them.
In particular, once one accepts our definition of fairness and finds a
set of fair solutions that satisfy all users, it becomes possible to use
the remaining freedom to select the specific solution that optimizes
some other metric.
For example, we can decide that the secondary goal is to maximize
system utilization;
in the above example, this will lead to preferring the symmetric
solution where all resources are bottlenecks over the other solutions
where only two are bottlenecks.
This provides an interesting way to combine user-centric metrics (the
entitlements) with system-centric metrics (resource utilization).

Of course, making such optimizations hinges on our ability to identify
and characterize all the possible solutions.
At present how to do this remains an open question.

\section{Conclusions}

To summarize, our main contribution is the definition of what it means
to make a fair allocation of multiple continuously-divisible resources
when users have different requirements for the resources, and a proof
that such an allocation is in fact achievable.
The definition is based on the identification of bottleneck resources,
and the allocation guarantees that each user either receives all he
wishes for, or else gets at least his entitlement on some bottleneck
resource.
The proof is constructive in the sense that it describes a method to
find such a solution numerically.
The method has in fact been programmed in Matlab, and was used in our
exploration of various scenarios.
While this method has seemed efficient in practice, 
one obvious open question is whether we can get a method that is
polynomial in $N$ and $m$.  

Note that, in the context of on-line scheduling, we may not need to
find an explicit solution in advance.
Consider for example the RSVT scheduler described by Ben-Nun et
al.\ \cite{bennun10}.
This is a fair share scheduler that bases scheduling decisions on the
gap between what each user has consumed and what he was entitled to
receive.
To do so, the system keeps a global view of resource usage by the
different users.
If there is only one bottleneck in the system, this would be applied
to the bottleneck resource.
The question is what to do if there are multiple bottlenecks.
Our results indicate that the correct course of action is to
prioritize each process based on the \emph{minimal} gap on any of the
bottleneck devices, because this is where it is easiest to close the
gap and achieve the desired entitlement.
Once the user achieves his target allocation on any of the bottleneck
devices, he should not be promoted further.
This contradicts the intuition that when a user uses multiple
resources, his global priority should be determined by the one
where he is farthest behind.

It should also be noted that our proposal pertains to the policy
level, and only suggests the considerations that should be applied
when fair allocations are desired.
It can in principle be used with any available mechanism for actually
controlling resource allocation, for example, resource containers
\cite{banga99}.

A possible direction for additional work is to extend the model.
In particular, an interesting question is what to do when the relative
usage of different resources is not linearly related.
In such a case, we need to replace the user-based factors $x_i$ by
specific factors $x_{ij}$ for each user and resource.
This also opens the door for a game where users adjust their usage
profile in response to system allocations --- for example,
substituting computation for bandwidth by using compression --- and
the use of machine learning to predict performance and make
optimizations \cite{bitirgen08}.
Finally, we might consider approaches where users have specific
utilities associated with each resource.  

\hide{
\subsection*{Acknowledgments}

Danny Dolev is Incumbent of the Berthold Badler Chair in Computer
Science, and was supported in part by the Israeli Science
Foundation (ISF) Grant number 1685/07.
Dror Feitelson was supported by the Israel Science Foundation (grant no.\
28/09), and by an IBM faculty award.
Joseph Halpern was supported in part by NSF grants ITR-0325453,
IIS-0534064, IIS-0812045, and IIS-0911036, by AFOSR grants
FA9550-08-1-0438 and FA9550-09-1-0266, ARO grant W911NF-09-1-0281, and a
Fulbright Fellowship.
}

\bibliographystyle{myabbrv}
\bibliography{abbrv,par,misc,se}
\end{document}